\documentclass[aps,prb,twocolumn,groupedaddress]{revtex4-1}
\usepackage[utf8]{inputenc}
\usepackage{graphics}
\usepackage{graphicx}
\usepackage{epstopdf}
\graphicspath{./fig/}

\begin{document}

\title{In-gap band in the one-dimensional two-orbital Kanamori-Hubbard model with inter-orbital Coulomb interaction}

\author{N. Aucar Boidi$^1$}
\email[]{nair.aucar@cab.cnea.gov.ar}
\author{H. Fern\'andez Garc\'{\i}a$^1$}
\email[]{Current affiliation: University of Antwerp, 2020 Antwerp, Belgium}
\author{Y. N\'u\~nez-Fern\'andez$^2$}
\author{K. Hallberg$^1$}
\affiliation{$^1$ Centro At{\'o}mico Bariloche and Instituto Balseiro, CNEA and CONICET, 8400 Bariloche, Argentina}
\affiliation{$^2$ CEA-Grenoble, IRIG-PHELIQS, 38000 Grenoble, France}

\date{\today}

\begin{abstract}
We study the electronic spectral properties at zero temperature of the one-dimensional (1D) version of the degenerate two-orbital Kanamori Hubbard model (KHM), one of the well established frameworks to study transition metal compounds, using state-of-the-art numerical techniques based on the Density Matrix Renormalization Group. While the system is Mott insulating for the half-filled case, as expected for an interacting 1D system, we find interesting and rich structures in the single-particle density of states (DOS) for the hole-doped system. In particular, we find the existence of in-gap states 
which are pulled down to lower energies from the upper Hubbard band (UHB) with increasing the inter-orbital Coulomb interaction $V$. We analyze the composition of the DOS by projecting it onto different local excitations and we observe that for large dopings these in-gap excitations are formed mainly by inter-orbital holon-doublon (HD) states and their energies follow approximately the HD states in the atomic limit. 
We observe that the Hund interaction $J$ increases the width of the in-gap band, as expected from the two-particle fluctuations in the Hamiltonian. 
The observation of a finite density of states within the gap between the Hubbard bands for this extended 1D model indicates that these systems present a rich excitation spectra which could help us understand the microscopic physics behind multi-orbital compounds.

\end{abstract}

\maketitle

\section{Introduction}

Understanding the microscopic mechanisms in materials with strong electron-electron correlations due to interactions in local orbitals, stands out as one of the most challenging problems in condensed matter physics. For example, materials like transition-metal oxides with partially filled $d$ or $f$ shells give rise to interesting properties like high temperature superconductivity, colossal magnetoresistance, correlation-driven metal-insulator transitions, heavy fermion behavior or an orbital selective Mott phase\cite{Vojta, Vojta2}. 

The discovery of a family of materials with similar characteristics in low dimensions, such as those with correlated electrons in ladders, gives us the possibility of a more detailed understanding of the underlying physical mechanisms. This is because of the availability of more accurate theoretical and numerical tools for one dimensional models, such as the Density Matrix Renormalization Group technique\cite{White, Karen1, Karen2, ramasesha, KW}. 
Among these systems we can mention the AFe$_2$X$_3$ family, where A: K, Rb, Cs, and Ba, and X are chalcogenides X: S, Se, which are formed by double chains of edge sharing FeX$_4$ tetrahedra.\cite{Takubo} In particular BaFe$_2$S$_3$\cite{Yamauchi, Takahashi} and BaFe$_2$Se$_3$\cite{Ying, Zhang} were found to superconduct under high pressure. The latter compound behaves in a manner compatible with orbital selective Mott physics\cite{Caron, Rincon, Dong}.
However, in spite of the theoretical progress made in the calculation of static properties, it is still difficult to obtain precise and detailed theoretical electronic structure results to compare with experiments, like angular resolved photoemission (ARPES), inverse photo-emission experiments (IPE) or optical conductivity measurements.

In this paper we report on results obtained for the zero-temperature local density of states (DOS) for the model Hamiltonian which decribes these multi-orbital systems, the two-orbital Kanamori-Hubbard model (KHM) \cite{KanamoriHubbard,reviewGeorges} in one dimension, which includes a ferromagnetic Hund coupling $J$ between the orbitals, in the doped regime, with equal bandwidths, using the DMRG for the calculation of static and dynamical properties \cite{Karen2, ramasesha, KW}.

By carefully analyzing the local electronic density of states (DOS) for a large range of parameters and dopings, we find a well defined in-gap band for large enough values of the inter-orbital Coulomb interaction ($V$). This structure is additional to the well known upper and lower Hubbard bands (UHB, LHB). As we show in this work, for large dopings, this in-gap band is formed mainly by holon-doublon pairs.
These results are similar to the findings in Refs. \cite{yurielhd,yurieldop,Yashar} for the two-orbital KHM but on a Bethe lattice using DMFT where an additional peak (in the half-filled case) or an additional band (for the doped case) were observed in the presence of $V$. 
In our case, as we are considering an interacting one-dimensional system, which is insulating for the half-filled case, we must dope the system to study its metallic behavior, where this new band appears. 
Although previous work reported holon-doublon pairs in related models at higher energies\cite{fulde,wang,picos,karski}, and also as metastable states out of equilibrium\cite{prelovsek,rinconfeiguin}, the existence of an in-gap band, like the one we are presenting in this paper, was not reported before for this model.

\section{Model and Method}
We study the degenerate two-orbital Kanamori-Hubbard model in 1D:
\begin{equation}
H=\sum_{\left\langle ij\right\rangle \alpha\sigma}t_{\alpha}c_{i\alpha\sigma}^{\dagger}c_{j\alpha\sigma} - (\mu -\epsilon )\sum_{i}n_{i} + \sum_{i}{H}_{i}\mbox{,}\label{eq:KHM}
\end{equation}
where $\left\langle ij\right\rangle $ are nearest-neighbor sites on a chain, $\alpha=1,2$ are orbital indices, and $\sigma$ the spin index.
The creation and destruction operators are $c^{\dagger}$ and $c$, respectively and $n_i=\sum_{\alpha\sigma}c_{i\alpha\sigma}^{\dagger}c_{i\alpha\sigma}$ the on-site number operator. The nearest-neighbor hoppings for orbital 1 and 2 take the value $t_1=t_2=0.5$ which is taken as the unit of energy. No inter-orbital hybridization is considered.
Here $\mu$ is the chemical potential where $\mu=0$ leads to half-filled bands. This implies that the site energies must take the values $\epsilon=-U/2 - V + J/2$

The on-site interactions $H_i$ are:
\begin{equation}
\begin{array}{c}
{H_{i}}=U\sum_{\alpha}n_{i\alpha\uparrow}n_{i\alpha\downarrow}+\sum_{\sigma\sigma'}\left(V-J\delta_{\sigma\sigma'}\right)n_{i1\sigma}n_{i2\sigma'}-\\
-J\left(c_{i1\uparrow}^{\dagger}c_{i1\downarrow}c_{i2\downarrow}^{\dagger}c_{i2\uparrow}+c_{i2\uparrow}^{\dagger}c_{i2\downarrow}c_{i1\downarrow}^{\dagger}c_{i1\uparrow}\right)\\
-J\left(c_{i1\uparrow}^{\dagger}c_{i1\downarrow}^{\dagger}c_{i2\uparrow}c_{i2\downarrow}+c_{i2\uparrow}^{\dagger}c_{i2\downarrow}^{\dagger}c_{i1\uparrow}c_{i1\downarrow}\right)
\end{array}\label{eq:interaction}
\end{equation}
where $J>0$ is the local exchange Hund's coupling and $U$ ($V$) is the intra (inter)-orbital Coulomb repulsion between electrons (see Fig. 1).

\begin{figure}
 \includegraphics[scale=0.25]{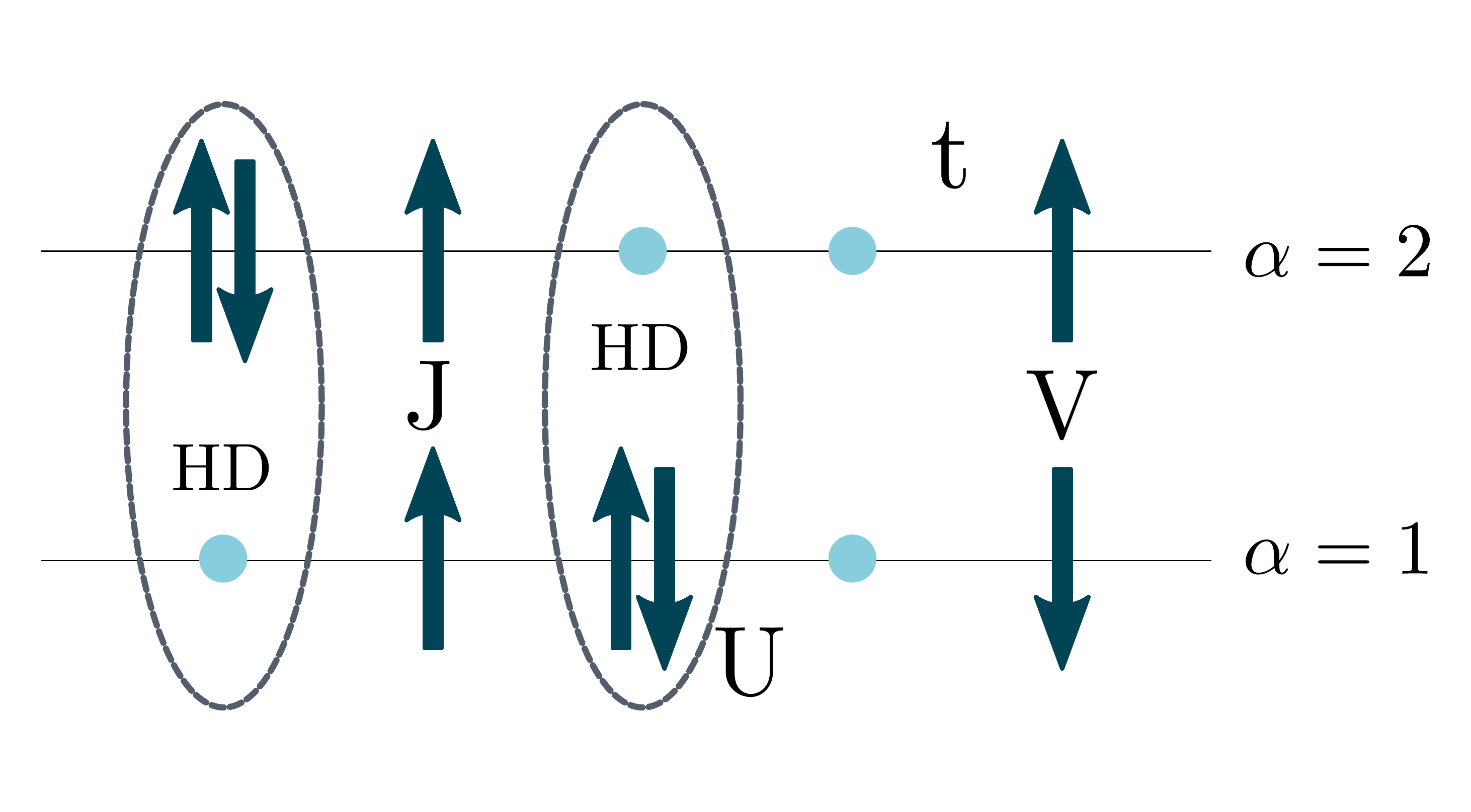}
  \caption{\label{figure1} Sketch of the interactions included in the KHM (Eq. (1)). Also shown is the representation of the inter-orbital HD excitation.}
 \end{figure}

 \subsection{Method}
To calculate the static and dynamical properties of the model we use the matrix-product-state (MPS) implementation of the DMRG in the OSP order (orbital-spin-position, varying the position index first) together with the inversion symmetry along the chain axis and open boundary conditions. 
However, for the calculation of the band dispersions in Fig. 7, we use periodic boundary conditions.
This geometry led to the most reliable results. 

For the DOS we calculate the following dynamical response functions:
\begin{equation}
A^{>}(\omega)=-\frac{1}{\pi}\Im\langle c_{1\uparrow}(\omega+i\eta-H+E_{0})^{-1}c_{1\uparrow}^{\dagger}\rangle\label{eq:Aq>} 
\end{equation}
\begin{equation}
A^{<}(\omega)=-\frac{1}{\pi}\Im\langle c_{1\uparrow}^{\dagger}(\omega+i\eta-H+E_{0})^{-1}c_{1\uparrow}\rangle\label{eq:Aq<}
\end{equation}
where the expectation is taken for the ground state with energy $E_{0}$ of the Hamiltonian (\ref{eq:KHM}).

All dynamical functions are obtained using the correction-vector method\cite{ramasesha} directly on the real axis (with a very small imaginary offset or Lorentzian broadening $\nu =0.15$) at zero temperature on a wide scale of energies. Smaller values of this broadening result in features in the DOS revealing its finite-size structure. The main results of this paper do not depend on this value.

We study system sizes of $L=12$ that correspond to $12$ physical lattice sites with two orbitals each. 
The number $m$ of states used to calculate ground states energies (the bond dimension) was set to 1024 and up to 512 for correlation functions and we performed up to eight sweeps reaching a good convergence with $m$.
We have also done calculations for other system sizes, showed in Fig. 6, to see possible size effects. In these results, we can observe that the main
features of the DOS remain the same. 

\section{Results}
  
\subsubsection{Zero Hund interaction $J=0$}
  
We will consider first the simpler case in which the Hund coupling is $J=0$ and study its effect later, including the rotationally invariant case defined when $V=U-2J$.\cite{reviewGeorges} 

For the half-filled case ($\mu=0$) the system is a Mott insulator for finite values of $U$ as seen in Fig. 2, where the UHB and LHB can be seen. 
When the system is doped with holes we observe an additional structure in between these Hubbard bands. This structure evolves towards higher energies with $\mu$, following the atomic HD excitations, marked with arrows (see Table I in Ref. \cite{yurieldop}). 
For $\mu \ge -U/2$ ($J=0$) the atomic limit HD excitation is $(U-V)$ while for $\mu < -U/2$ it is $(U/2-V-\mu)$. 
The light blue line corresponds to the projection of the DOS onto local HD excitations as explained in Sect. IV.

 \begin{figure}
 \includegraphics[scale=0.3]{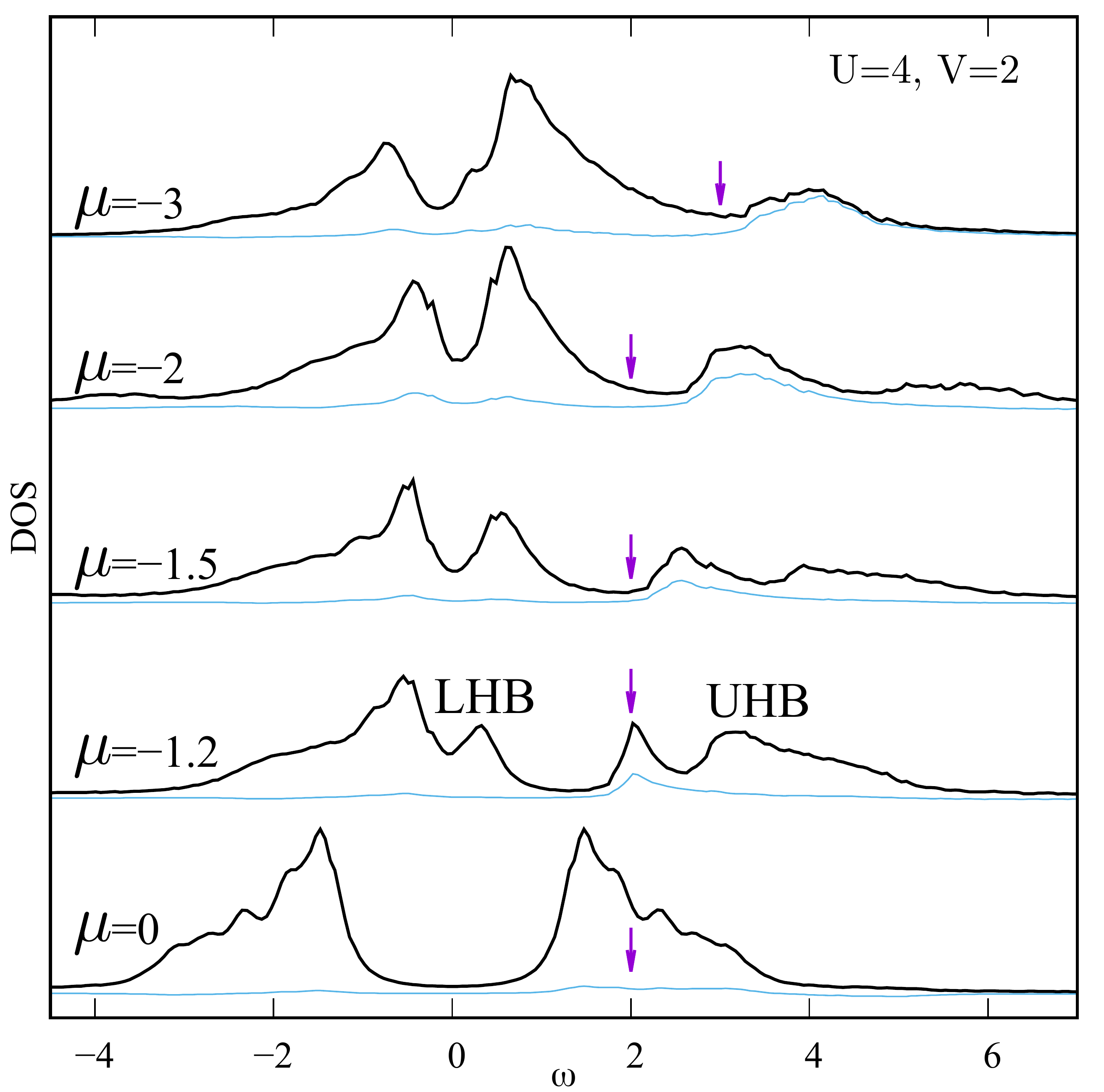}
  \caption{\label{figure2} DOS for the two orbital KHM for $U=4$, $J=0$, $L=12$, and $V=2$ while varying the chemical potential $\mu$. $\omega=0$ corresponds to the Fermi energy in all figures. For $\mu=0$ the UHB and LHB are present and the system is a Mott insulator. When the system is doped with holes an in-gap band is formed.
  Also shown is $A_{HD}(\omega)$ (light blue line) and the energies of the HD state in the atomic limit (arrows).}
  \end{figure}

It is also interesting to analyze the effect of the inter-orbital interaction $V$, as we show in Fig. 3. When $V=0$ the system consists of two independent interacting Hubbard chains and both Hubbard bands are clearly recognizeable. When $V$ is turned on we observe a transfer of weight from the UHB to an in-gap band which decreases in energy as $V$ grows. The case $V=4=U$ corresponds to the rotationally invariant situation ($J=0$) and the in-gap band merges with the central structure.
We also observe a dip in the DOS at the Fermi energy $\omega=0$ for finite values of $V$. As studied below (see Sect. IV), these states are formed mainly by excitations of the kind $|0,\uparrow\rangle\ $ and $|\uparrow,\uparrow\rangle $ (and all other spin projections) in the atomic limit.

\begin{figure}
 \includegraphics[scale=0.3]{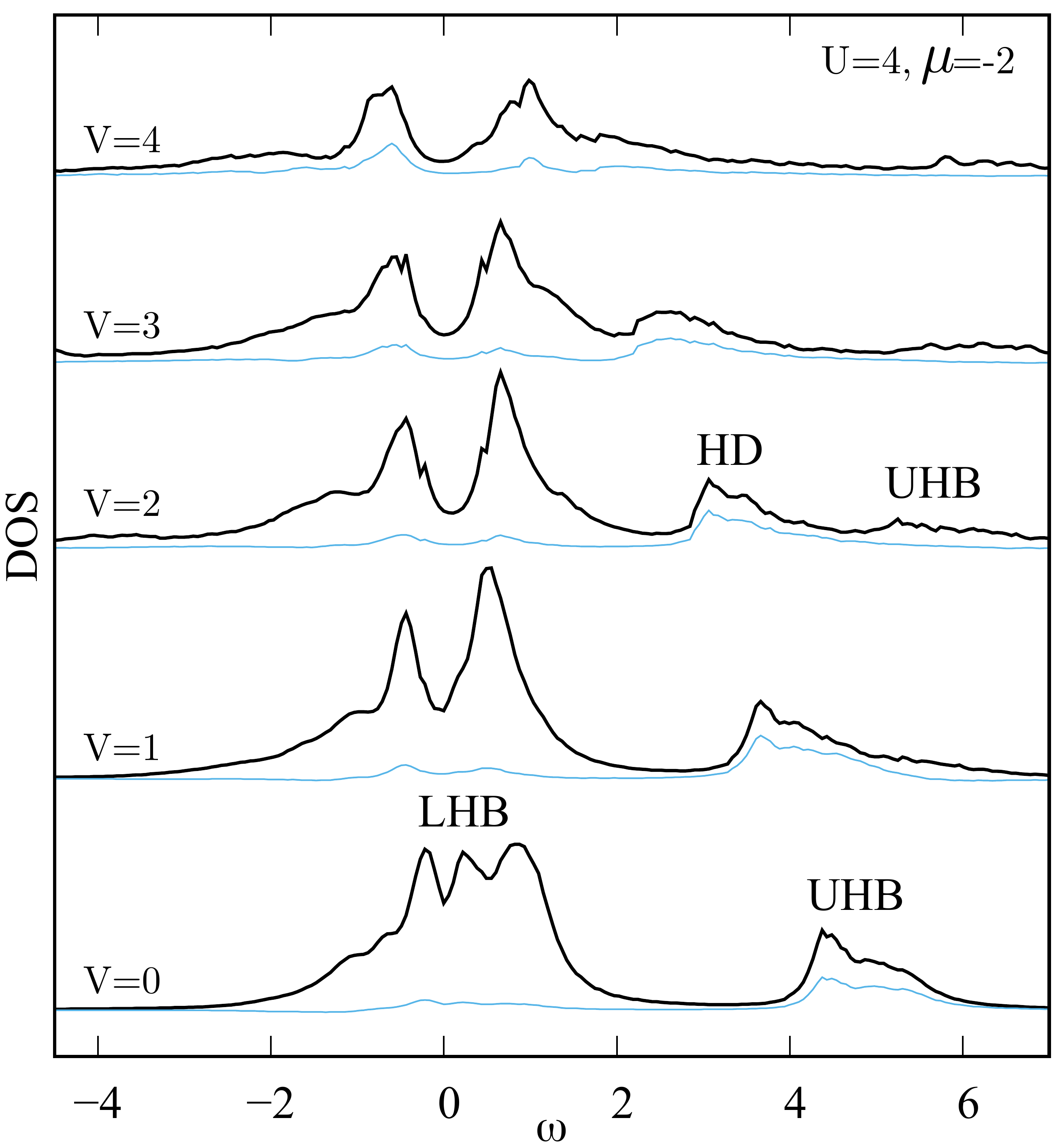}
  \caption{\label{figure3} DOS for the two orbital KHM for $U=4$, $J=0$, $L=12$, and $\mu=-2$ while increasing the inter-orbital interaction $V$. For $V=0$ the UHB and LHB are present and the system is a metal. Also shown is $A_{HD}(\omega)$ (light blue line).}
 \end{figure}

\subsubsection{Finite Hund interaction $J$}
In Fig. 4 we present results for the DOS in the presence of a finite inter-orbital Hund interaction $J$. We observe changes in the DOS at low energies and also in the in-gap band. As we show below, the DOS close to the Fermi energy $\omega=0$ is formed, mainly, by singly occupied stated in both orbitals, which are affected by the spin flip term in the Hamiltonian (\ref{eq:KHM}). The widening produced by $J$ in the in-gap band is produced, on the other hand, by the two-particle fluctuations in \ref{eq:KHM} since this band is formed mainly by HD pairs.\cite{conJ}

\begin{figure}
 \includegraphics[scale=0.3]{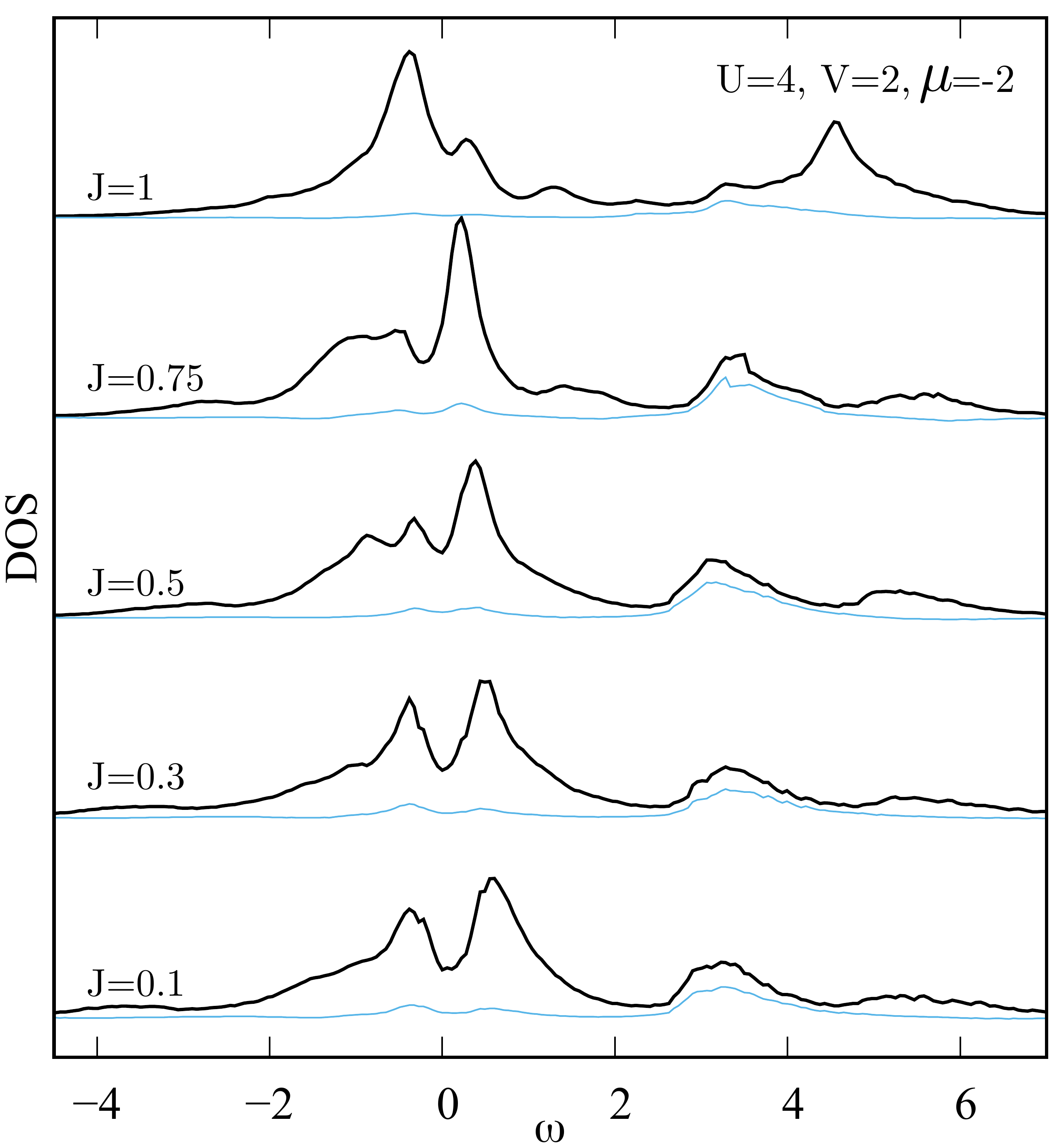}
  \caption{\label{figure4} DOS for the two orbital KHM for $U=4$, $V=2$, $L=12$, and $\mu=-2$ while increasing the inter-orbital Hund interaction $J$. $J=1$ corresponds to the rotationally symmetric case $V=U-2J$. Also shown is $A_{HD}(\omega)$ (light blue line).}
 \end{figure}

\section{Characterization of the excitations}

In order to understand the effect of the inter-orbital interaction $V$ on the DOS we have calculated several dynamical response functions which correspond to the projection of the DOS onto particular atomic states on each site: one site is composed by two orbitals and its states is represented as $|s1,s2\rangle$. 
In a similar way as done in Ref. \cite{yurieldop} we define the Green's functions $A_{s_{1},s_{2}}(\omega)=A_{s_{1},s_{2}}^{>}(\omega)+A_{s_{1},s_{2}}^{<}(-\omega)$
with:
\begin{equation}
A_{s_{1},s_{2}}^{>}(\omega)=-\frac{1}{\pi}\Im\langle c_{1\uparrow}(\omega+i\eta-H+E_{0})^{-1}X_{s_{1},s_{2}}^{\dagger}\rangle\label{eq:Aq>} 
\end{equation}
\begin{equation}
A_{s_{1},s_{2}}^{<}(\omega)=-\frac{1}{\pi}\Im\langle c_{1\uparrow}^{\dagger}(\omega+i\eta-H+E_{0})^{-1}X_{s_{1},s_{2}}\rangle\label{eq:Aq<}
\end{equation}
where the expectation is taken for the ground state with energy $E_{0}$ of the Hamiltonian (\ref{eq:KHM}).
The excitations are $X_{s_{1},s_{2}}^{\dagger}=P_{s_{1},s_{2}}c_{1\uparrow}^{\dagger}$
and their reverse action $X_{s_{1},s_{2}}=c_{1\uparrow}P_{s_{1},s_{2}}$.
The projector $P_{s_{1},s_{2}}=|s_{1},s_{2}\rangle\langle s_{1},s_{2}|$
is used to select the corresponding atomic configuration $|s_{1},s_{2}\rangle$.
Note that adding all possible configurations gives the total DOS since 
$\sum_{s_{1},s_{2}}P_{s_{1},s_{2}}=1$.

We are particularly interested in the following excitations (and their
reverse actions) for orbital 1 (similarly for orbital 2), where we add over all spin projections:

\noindent (i)  HD states ($|\downarrow,0\rangle\to|\uparrow\downarrow,0\rangle=|d,0\rangle$):
$X_{d,0}^{\dagger}=n_{1\downarrow}(1-n_{2\uparrow})(1-n_{2\downarrow})c_{1\uparrow}^{\dagger}$

\noindent (ii)  $|0,0\rangle\to|\uparrow,0\rangle$:
$X_{\uparrow,0}^{\dagger}=(1-n_{1\downarrow})(1-n_{2\uparrow})(1-n_{2\downarrow})c_{1\uparrow}^{\dagger}$

\noindent (iii)  $|0,\uparrow\rangle\to|\uparrow,\uparrow\rangle$:
$X_{\uparrow,\uparrow}^{\dagger}=n_{2\uparrow}(1-n_{2\downarrow})(1-n_{1\downarrow})c_{1\uparrow}^{\dagger}$

\noindent (iv)  $|\downarrow,\uparrow\rangle\to|\uparrow\downarrow,\uparrow\rangle=|d,\uparrow\rangle$:
$X_{d,\uparrow}^{\dagger}=n_{2\uparrow}(1-n_{2\downarrow})n_{1\downarrow}c_{1\uparrow}^{\dagger}$ \\

In Fig. 5 we present the results for the projection of the DOS onto these atomic states. It is interesting to observe that for $V=0$ (Fig. 5a) we do not observe the in-gap structure and all excitations containing doublons contribute to the UHB, as expected. When we turn on the inter-orbital Coulomb interaction $V$ (Fig. 5b) there is a transfer of spectral weight towards lower energies forming a new band which is constituted mainly by HD excitations. The weight of these HD excitations in the UHB is negligible. When the doping is increased further (e.g. $\mu=-3$, (Fig. 5c)) the HD character of the new band is increased and we find, as expected, a reduction in the total weight of the UHB.
The curve corresponding to the HD projection is also shown in the previous figures to signal the extra band. In Fig. 2 we see how the HD excitations increase their energy with hole doping (also marked with an arrow following the projection onto the atomic HD states as previously explained). Here we find that for larger dopings the in-gap band has a larger weight in the HD states.
In Fig. 3 we see how this HD projection accompanies the new band towards lower energies with $V$ while in Fig. 4 we see how this new band widens slightly with $J$.

We also find that the low energy excitations are formed mainly by states of the form $ |0,\uparrow\rangle\ $ and $ |\uparrow,\uparrow\rangle $. The observation of regions with charge density waves (CDW) and superconductivity for the low hole doping limit of the rotationally invariant case of this model in Ref. \cite{Dagotto} suggests that this low energy structure could be related to superconducting or CDW fluctuations. We leave the analysis of these low energy excitations for further study.

\begin{figure}
 \includegraphics[scale=0.3]{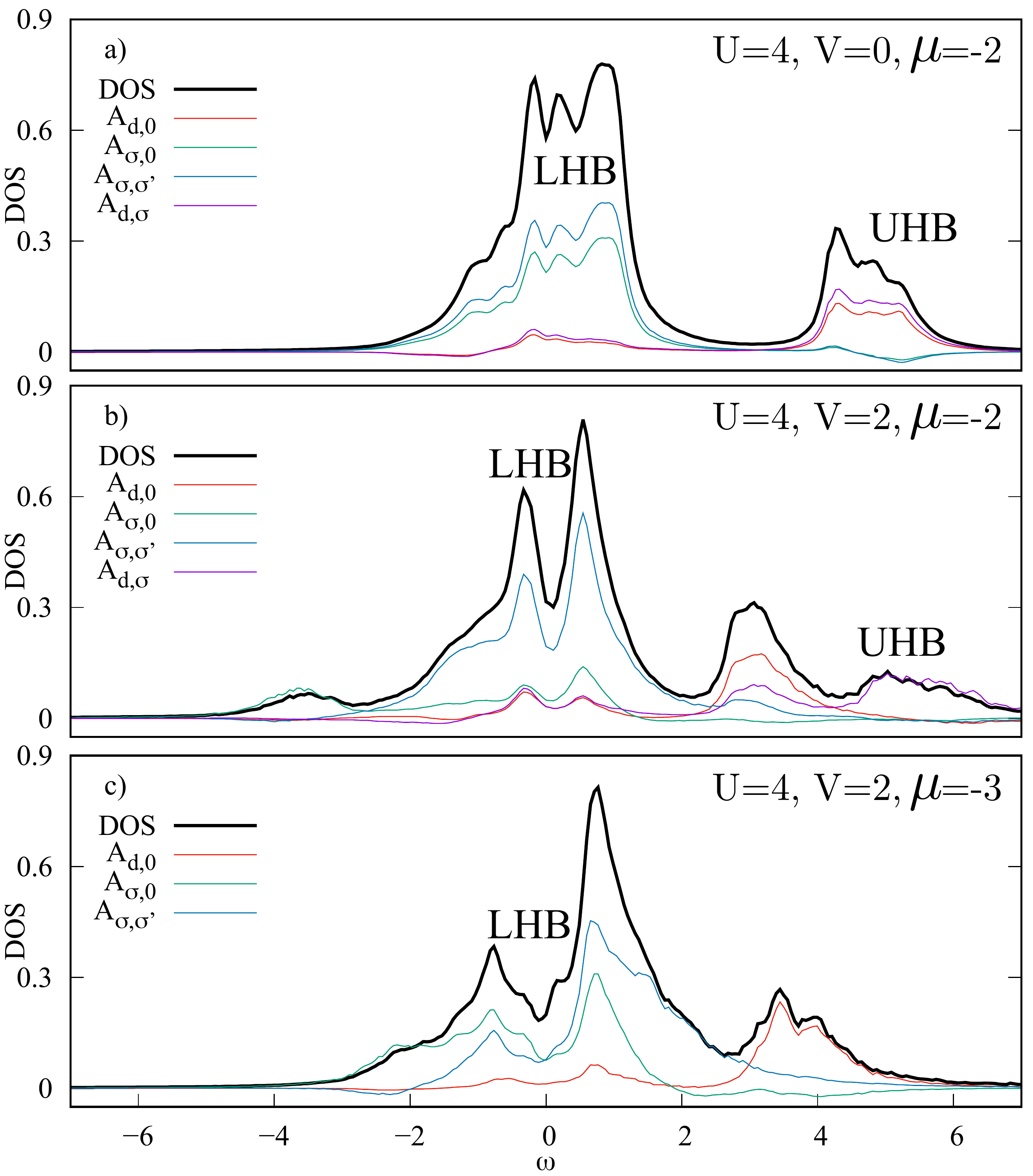}
  \caption{\label{figure5} DOS projected onto different atomic states defined in the text for $U=4$, $J=0$ and $L=12$. a) $V=0$, $\mu=-2$; 
  b) $V=2$, $\mu=-2$ and c) $V=2$, $\mu=-3$ (other projections have negligible weight and are not shown).}
 \end{figure}

In Fig. 6 we plot the DOS for several system sizes, showing that the existence of the in-gap band is robust. 

\begin{figure}
 \includegraphics[scale=0.3]{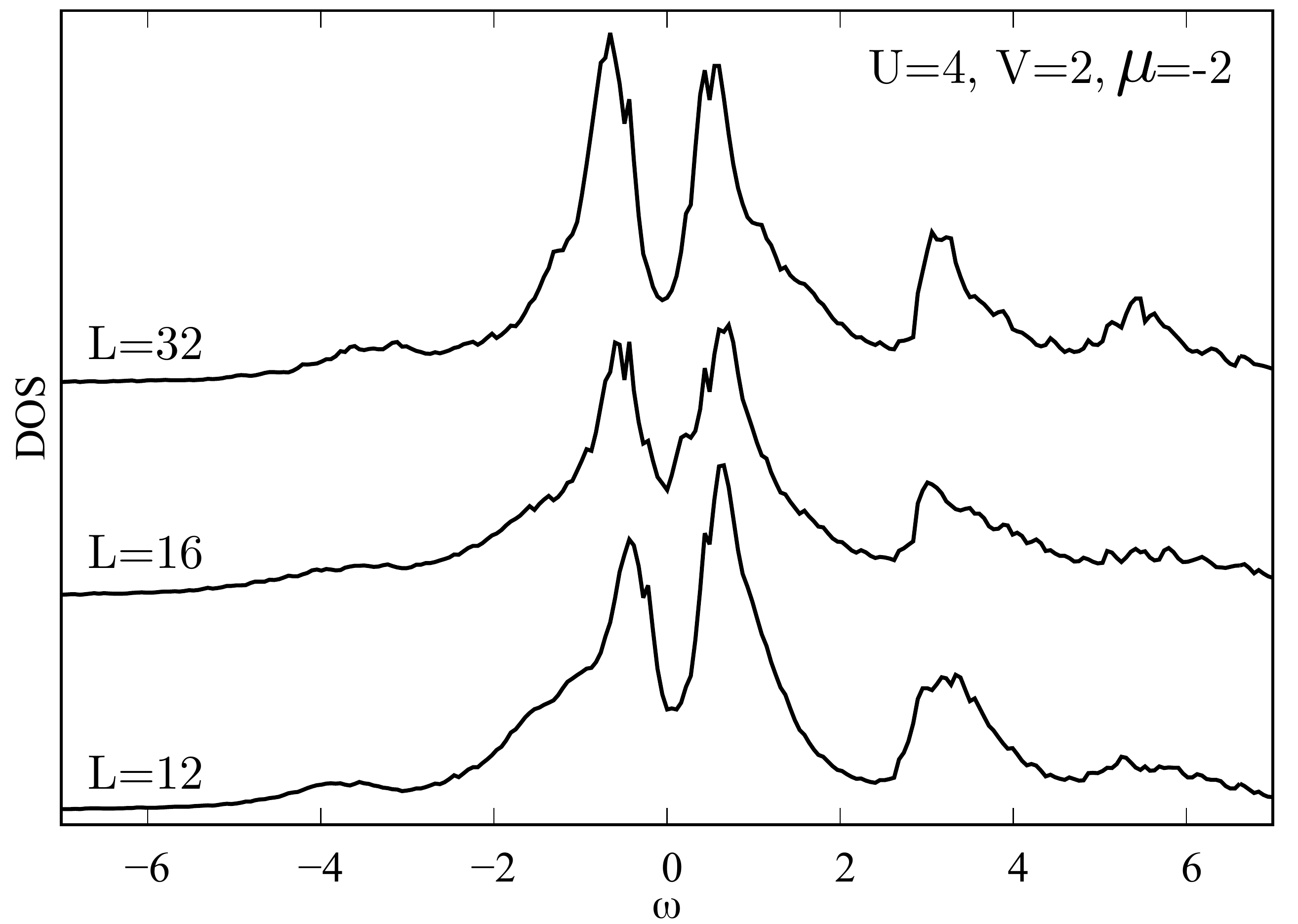}
  \caption{\label{figure6} DOS for different system sizes ($L$ is the number of two-orbital sites) for $U=4$, $V=2$, $J=0$ and $\mu=-2$.}
 \end{figure}

\section{Band dispersions}
We also calculate the spectral functions $A(\omega,k)$, which are obtained by Fourier transforming the single-particle Green's function in real space with periodic boundary conditions using DMRG.

The momentum dispersions of the bands are plotted in Fig. 7. In the right column we show the dispersion of the HD excitations. For $V=0$ (which corresponds to two independent single-orbital chains) we observe that the HD excitations are located mainly at the zone boundaries, $k=\pm \pi $, while for finite $V$ the HD excitations seem to be more extended in $k$. In this latter case we can clearly see that the HD band is formed at lower energies, as shown before with the DOS. We also see the dip in the DOS at low energies. For $J=1$ we observe a broadening of the HD band together with a reduction of the dip close to the Fermi energy ($\omega =0$).

\begin{figure}
\includegraphics[scale=0.35]{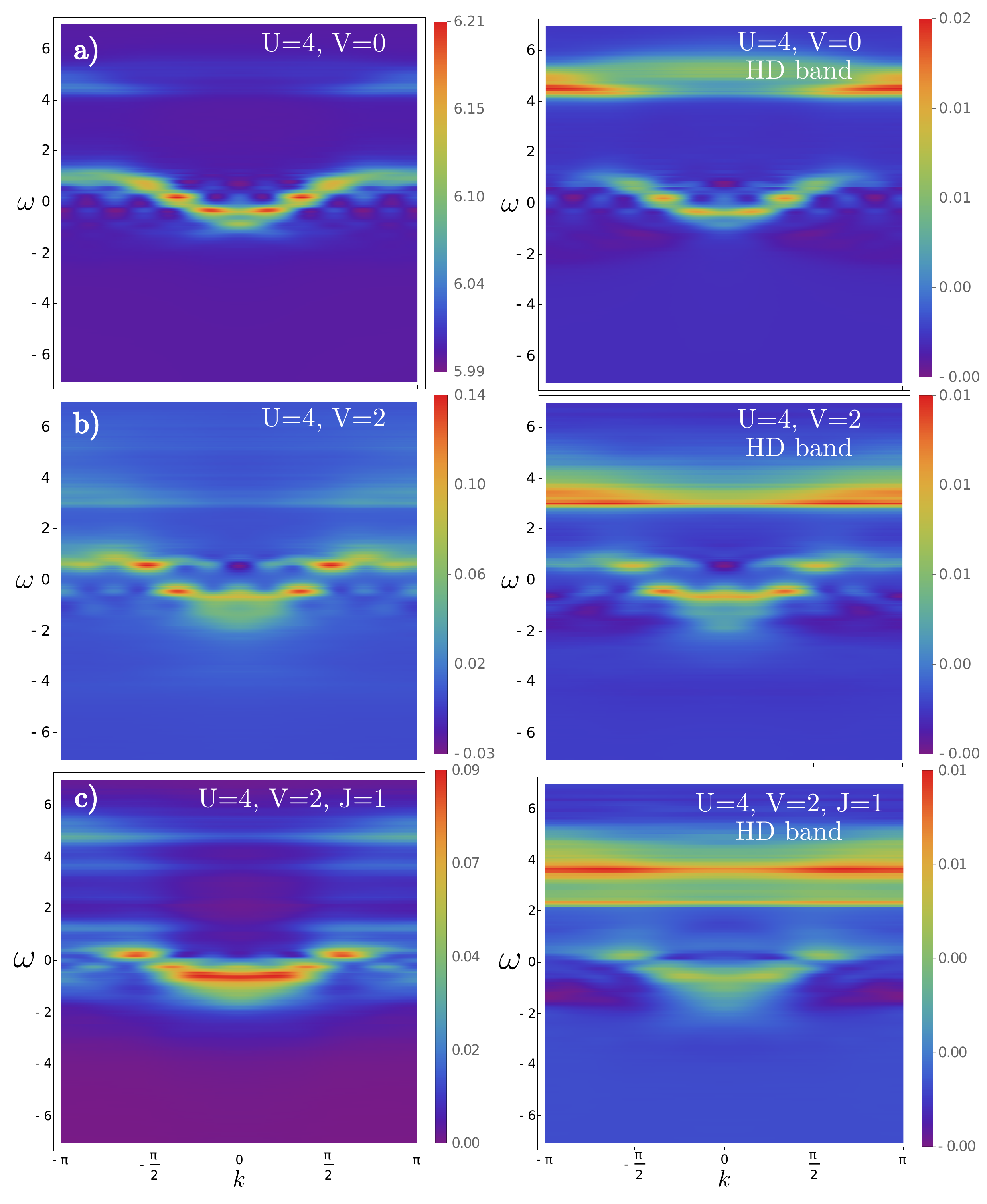}
  \caption{\label{figure7} Heat plots of $A(\omega,k)$ for $U=4$, $\mu=-2$ and $L=12$ and periodic boundary conditions. a) $V=0$,
$J=0$, b) $V=2$, $J=0$ and c) $V=2$, $J=1$ (rotationally symmetric case).
Figures in the right column show the dispersions projected onto de HD excitations for each case.}
 \end{figure}

\section{Conclusions}

In this work we studied the $T=0$ single electron spectral functions of the 1D two-orbital Kanamori-Hubbard model which is the basic model to describe a wide range of correlated multi-orbital materials. 
We resort to high precision DMRG numerical calculation for the ground states and the dynamical response functions. 
This allowed us to observe a rich structure in the DOS for a wide range of parameters. A salient feature for the hole doped case is the presence of an in-gap band with a large component on holon-doublon excitations whose spectral weight is transfered from the upper Hubbard band for intermediate values of the inter-orbital Coulomb interaction $V$. We expect that these in-gap excitations will be also observed in two and three dimensional versions of this model, as already indicated in previous work using DMFT. \cite{yurielhd,yurieldop} One main difference is that for higher dimensions one can also analyze the half-filled case which is metallic for small enough local interactions. As shown in Ref. \cite{yurielhd}, the HD band is also observed in this case, where a metallic band is needed to provide for the holes or the doubly occupied states which form the HD excitations. For the one-dimensional case studied in this paper, the holes are provided by doping.

We also studied the energy dispersion and found that these excitations are less concentrated close to the zone boundaries when compared to the $V=0$ case. The new in-gap band shown in this paper should also show up in other observables such as optical conductivity. 
Work is underway to analyze the low energy features and the consequences of electron doping.

\begin{acknowledgments}
We acknowledge support from projects PICT 2016-0402 and PICT 2018-01546 from the Argentine ANPCyT and PIP2015 - 11220150100538 CO (CONICET). We thank Cristian Batista, Adriana Moreo and Elbio Dagotto for useful dicussions.
\end{acknowledgments}


\end{document}